\documentclass[12pt,a4paper,twoside]{article}
\usepackage{t1enc}
\usepackage[latin1]{inputenc}
\usepackage[english]{babel}
\usepackage{yfonts}
\usepackage{bbm}
\usepackage{bm}
\usepackage{amssymb}
\usepackage{amsmath}
\usepackage{amsfonts}
\usepackage{fancyhdr}
\usepackage[all,cmtip,line]{xy}

\makeatletter
\renewcommand{\section}{\@startsection{section}{1}{0mm}{-\baselineskip}{.25\baselineskip}{\normalfont \Large \scshape}}
\makeatother

\makeatletter
\renewcommand{\subsection}{\@startsection{subsection}{2}{0mm}{-\baselineskip}{.25\baselineskip}{\normalfont \large \scshape}}
\makeatother

\renewenvironment{abstract}[0]
{\begin{center} {\scshape Abstract} \end{center} \par \vspace{-.6cm} \begin{center} \small \begin{minipage}[t]{.9\textwidth}}
{ \end{minipage}\end{center} \vspace{.3cm} }

\newcommand{\bra}[1]{\langle #1|}
\newcommand{\ket}[1]{| \, #1\rangle}

\begin{document}

\scshape\title{\textsc{Quantization of Wilson loops in Wess-Zumino-Witten models}}
\author{\\ \textsc{Anton Alekseev\footnote{anton.alekseev@math.unige.ch}  \hspace{.2cm} \& \hspace{.2cm}  Samuel Monnier\footnote{samuel.monnier@math.unige.ch}} \\ \\ \small Université de Genève, Section de Mathématiques, \\ \small 2-4 Rue du Lièvre, 1211 Genève 24, Switzerland}
\maketitle

\upshape

\begin{abstract}
We describe a non-perturbative quantization of classical Wilson loops
in the WZW model. The quantized Wilson loop is an operator acting on the Hilbert space of closed strings and commuting either with the full Kac-Moody chiral algebra or with one of its 
subalgebras. We prove that under open/closed string duality, it is dual to a boundary perturbation of the open string theory. As an application, we show that such operators are useful 
tools for identifying fixed points of the boundary renormalization group flow.
\end{abstract}

\section{Introduction}

In this note, we propose a solution to the quantization of classical Wilson loops in the WZW models, a problem posed in \cite{BG}.

Wess-Zumino-Witten models are Lie group valued conformal field theories, and they describe the dynamics of strings in group manifolds. Because they are exactly solvable, they provide interesting models for string backgrounds with non-trivial metric and B-field. They also find applications in solid state physics, for instance in the Kondo problem \cite{AFF}.

Defining such a conformal field theory on a surface with boundaries requires to specify boundary conditions (D-branes). The conformal symmetries of the theory, as well as the expected modular properties of the correlation functions impose severe restrictions on the set of admissible boundary conditions \cite{Cardy:1989ir, Behrend:1999bn}. Because of the factorization properties of CFTs, one needs to provide the one point correlators for the bulk fields on the disk to fully describe a brane. Hence, by the state-operator mapping, branes can be seen as functionals on the state space of the model. By a slight abuse, one pictures them as coherent (non-normalisable) states in the state space, the ``boundary states''.

\vspace{.5cm}

Consider a compact worldsheet, and let $C$ be a cycle in the worldsheet. One can build the following Wilson loop :
\begin{equation}
\label{WloopIntro}
W^{(\frac{1}{k},\mu)}(J) = \mbox{Tr} \, \mbox{P} \exp \left ( \frac{i}{k} \int_C dz J^a(z) A^a \right ) \;,
\end{equation}
where P denotes cyclic ordering along $C$, $J^a(z)$ are the currents of the model and $A^a$ is a set of matrices forming a representation of the horizontal finite Lie algebra. The operator displayed above only has a precise meaning in the classical case, when $J^a(z)$ are complex valued functions on the worldsheet. In this setting it computes the monodromy of the chiral part of the classical solution of the WZW model (see \cite{BG}, §3). 

In \cite{BG}, the authors considered formal perturbations of boundary states of the form :
\begin{equation}
\label{PertClosed}
n \ket{B} \rightarrow W^{(\frac{1}{k},\mu)}(J) \ket{B} \;,
\end{equation}
where $n$ is the dimension of the representation. The reason why this is only a formal expression is that at the quantum level, $J^a(z)$ become Kac-Moody currents. As a result, the Wilson loop as defined above suffers from ordering ambiguities, and one quickly run into infinities when trying to expand it explicitly. \cite{BG} renormalized it up to order four in $\frac{1}{k}$, but computations quickly become intricate and it is not clear whether this procedure can be sucessful at any order. 

The non-perturbative quantization procedure developed below will allow us to associate a well-defined operator on the Hilbert space of the model to the classical Wilson loops (\ref{WloopIntro}). We will also show that the perturbation (\ref{PertClosed}) in the closed string picture is dual to some familiar boundary perturbations \cite{ARS} in the open string picture. We will integrate these perturbations explicitly for a special (finite) value of the coupling and show that they give rise to a WZW model with a new brane at the boundary supporting the perturbation. This brane is exactly the one given by the action of the quantized Wilson operator on the original boundary state. Hence it turns boundary perturbation into an efficient tool for finding new branes, allowing to compute the resulting boundary state in a straightforward manner.

We will also use this procedure in a concrete example taken from \cite{Monnier:2005jt}, in which some interesting branes were characterized only by the boundary perturbation generating them. We will find the boundary states explicitly, identifying them with the ones described in \cite{Quella:2002ct}.

Wilson loops are instances of conformal defects, which have been studied from various points of view both in conformal field theories \cite{Bazhanov:1994ft} and 3D topological field theories \cite{Felder:1999cv}. The idea of inserting central operators in correlation functions was introduced (to our knowledge) in \cite{Petkova:2000ip}, and \cite{Graham:2003nc} let them act on boundary states to make predictions about the boundary RG flow. The interest of our quantization scheme is that it allows to relate in a precise and rigorous manner boundary perturbations to such operators.

\vspace{.3cm}

In section \ref{bcft} we quickly review boundary conformal field theory of WZW models, introduce Wilson loops and explain the quantization problem. In section \ref{Kac}, we review \cite{Kac}, which contains key results for the quantization procedure. We solve the quantization problem in section \ref{quant}, and prove their relation with boundary perturbation of open string theories in section \ref{BPert&WLOp}. Finally, we show the power and simplicity of this formalism with an application to renormalization flows in section \ref{app}.

\section{Boundary conformal field theory}

\label{bcft}

Here we review some basic facts about boundary conformal field theory of WZW models 
(see \cite{Zuber:2000ia} and \cite{Gaberdiel:2002my} for pedagogical reviews of
boundary CFT). Consider a WZW model at level $k$ on a compact simple Lie group $G$. 
Let $\textfrak{g} = \mbox{Lie}(G)$ be the corresponding simple Lie algebra, and 
$\hat{\textfrak{g}}$ the associated affine Kac-Moody algebra at level $k$. 
The full symmetry of the model is generated by two commuting copies of $\hat{\textfrak{g}}$, 
with currents $J(z)$ and $\bar{J}(\bar{z})$. We will only consider models with 
diagonal partition functions, therefore the Hilbert space decomposes as 
$\mathcal{H} = \bigoplus_{\mu \in P^+_k} H_\mu \otimes H_{\mu^*}$. Here $P^+_k$
denotes the set of integral dominant weights of $\hat{\textfrak{g}}$ at level $k$, 
and $H_\mu$ the irreducible $\hat{\textfrak{g}}$-module of highest weight $\mu$.

\subsection{Maximally symmetric boundary conditions}

Maximally symmetric boundary conditions preserve the diagonal part of 
$\hat{\textfrak{g}} \otimes \hat{\textfrak{g}}$, the Kac-Moody algebra of the bulk theory. 
When the theory is quantized on a cylinder, the corresponding boundary state 
$|B \rangle$ satisfies the gluing condition :
\begin{equation}
\label{symcond}
(J_n + \bar{J}_{-n}) |B \rangle = 0 \;,
\end{equation}
where $J_n$ and $\bar{J}_n$ are the Fourier modes of the Kac-Moody currents. 
The Ishibashi states $\{| \: \mu \rangle \! \rangle \}$, $\mu \in P^+_k$ generate 
the space of all solutions to this equation. Their normalization can be fixed by the 
following overlap condition \nolinebreak:
$$
\langle \! \langle \lambda | \: q^{\frac{1}{2}(L_0 + \bar{L}_0 - \frac{c}{12})} | \: \mu \rangle \! \rangle = \delta_{\mu\lambda}\chi^{\hat{\textfrak{g}}}_\mu(q) \;\;\;\; q = e^{2\pi i\tau} \;,
$$
where $\tau$ is the modular parameter of the cylinder, $L_0$ and $\bar{L}_0$ are the zero modes of the holomorphic and antiholomorphic components of the Sugawara energy-momentum tensor, and $\chi^{\hat{\textfrak{g}}}_\mu(q)$ is the character of the Kac-Moody algebra $\hat{\textfrak{g}}$ evaluated at the weight $-2\pi i \tau \hat{\omega}_0$. 

Not all linear combinations of Ishibashi states provide consistent boundary states, 
the latter have to satisfy a supplementary condition, known as Cardy condition (\cite{Cardy:1989ir}).
Such a condition is necessary to ensure the modular invariance of the boundary conformal field theory,
and requires that maximally symmetric boundary states are linear combinations with positive integral 
coefficients of states of the following form \nolinebreak:
$$
|B_\mu\rangle = \sum_{\lambda \in P^+_k} \frac{S_{\mu\lambda}}{\sqrt{S_{0\lambda}}} |\lambda \rangle\!\rangle \;,
$$
where $S_{\rho\sigma}$ ($\rho,\sigma\in P^+_k$) are the elements of the modular $S$ matrix of the WZW model, 
and 0 is the zero weight. These states are called ``elementary'' maximally symmetric boundary states.

\subsection{Symmetry breaking boundary conditions}

\label{SymBreakBC}

The states introduced above preserve half of the bulk symmetry of the WZW model. However, the model
only needs to be invariant under conformal transformations leaving the boundary fixed. Hence it is 
natural to try to build boundary states which have less symmetry \cite{Fuchs:1999zi}, and in particular states which preserve only an affine subalgebra of $\hat{\textfrak{g}}$ \cite{Quella:2002ct}. We will summarize here part of the results appearing in \cite{Quella:2002ct}.

In \cite{Quella:2002ct}, the authors chose a semi-simple subgroup $A$ of $G$, and considered branes
preserving the symmetry algebra $\hat{\textfrak{a}} \oplus \hat{\textfrak{c}}$, where $\hat{\textfrak{a}}$
is the affine Kac-Moody algebra associated to $A$, and $\hat{\textfrak{c}}$ is the chiral algebra of 
the coset model $G/A$.

The Ishibashi states $|\, \mu, \sigma \rangle \! \rangle$ are labeled by the sectors of the coset model,
i.e. by pairs $(\mu, \sigma)$ of weights of $\textfrak{g}$ and $\textfrak{a}$, respectively. Not all of these pairs are admissible, the weights have to satisfy the algebraic relation : $\pi(\mu) - \sigma \in \pi(Q)$, where $\pi$ is the projection associated with the embedding $\textfrak{a} \subset \textfrak{g}$, and $Q$ is the weight lattice of $\textfrak{g}$. Moreover, some pairs label the same coset sector, and hence have to be identified (see \cite{Quella:2002ct} for details). The resulting set $P_{G/A}$ of pairs labels representations of the coset symmetry algebra as well as the Ishibashi states. The Ishibashi states are normalized as :
$$
\langle \! \langle \lambda, \sigma | \: q^{\frac{1}{2}(L_0 + \bar{L}_0 - \frac{c}{12})} | \, \mu, \zeta \rangle \! \rangle = \delta_{\lambda\mu}\delta_{\sigma\zeta} \, \chi^{\hat{\textfrak{c}}}_{(\mu,\zeta)}(q) \, \chi^{\hat{\textfrak{a}}}_\zeta(q) \;,
$$
where the coset characters $\chi^{\hat{\textfrak{c}}}_{(\mu,\zeta)}(q)$ satisfy the branching relation :
$$
\chi^{\hat{\textfrak{g}}}_\mu(q) = \!\!\!\! \sum_{\zeta \, | \, (\mu,\zeta) \in P_{G/A}} \!\!\!\! \chi^{\hat{\textfrak{c}}}_{(\mu,\zeta)}(q) \, \chi^{\hat{\textfrak{a}}}_\zeta(q) \;.
$$
Again, imposing Cardy's condition yields the following elementary boundary states :
\begin{equation}
\label{sbst}
|B_{(\lambda, \sigma)} \rangle = \!\!\!\! \sum_{(\mu, \zeta) \in P_{G/A}} \! \frac{S_{\lambda\mu}\bar{S}^{\hat{a}}_{\sigma\zeta}}{\sqrt{S_{0\mu}}\bar{S}^{\hat{a}}_{0\zeta}} |\mu, \zeta \rangle \! \rangle \;,
\end{equation}
where $S^{\hat{a}}$ is the modular $S$-matrix associated with the subalgebra $\hat{\textfrak{a}}$. Physical boundary states preserving $\hat{\textfrak{a}} \oplus \hat{\textfrak{c}}$ are linear combinations of these states with positive integer coefficients.

\subsection{Wilson loops}

\label{SectWloop}

Consider a classical WZW model at level $k$. This model is defined by a map $g(z,\bar{z})$ from the worldsheet to some compact simple Lie group $G$, and the action of the model is invariant when $g$ is multiplied from the left by a holomorphic $G$-valued function, and from the right by an antiholomorphic $G$-valued function. The Lie algebra corresponding to this group of symmetries is generated by the holomorphic and antiholomorphic currents \nolinebreak:
$$
J(z) = -ik\partial g g^{-1} \;\;\;\;\;\;\;\;\;\;\;\;\;\;\;\;  \bar{J}(\bar{z}) = ikg^{-1}\partial g \;.
$$
From the holomorphic current, we can build the following Wilson loop :
\begin{equation}
\label{WLoop}
W^{(l,\mu)}(J) = \mbox{Tr}_{R_\mu} \mbox{P} \exp \left ( i l \int_C dz J^a(z) A^a \right ),
\end{equation}
where $l \in \mathbb{R}$, $\mu$ is a weight of $\textfrak{g} = \mbox{Lie}(G)$, $A^a$ are matrices of the generators of $\textfrak{g}$ in the corresponding representation $R_\mu$, $C$ is a loop in the worldsheet, and $P$ stands for cyclic ordering along $C$. These classical operators are topological, which means that they depend only on the homotopy class of the loop $C$. At the special coupling $l = \frac{1}{k}$, the Wilson loop is even invariant under the full WZW symmetry group (it has vanishing Poisson bracket with the currents $J$ and $\bar{J}$, see \cite{BG}).

The goal is to quantize the classical Wilson loop at $l = \frac{1}{k}$. Its quantum counterpart will be an operator acting on the Hilbert space of the quantum theory, namely on integrable highest weight modules of the Kac-Moody algebra. As the classical loop is gauge invariant, we will demand that the quantum operator should commute with the full symmetry algebra $\hat{\textfrak{g}} \otimes \hat{\textfrak{g}}$. Being independent of $\bar{J}$, it automatically commutes with the antiholomorphic copy of $\hat{\textfrak{g}}$. To commute with the holomorphic chiral algebra, it should be \emph{central} with respect to the holomorphic copy of $\hat{\textfrak{g}}$. More precisely, it should belong to the center of some completion of the enveloping algebra of $\hat{\textfrak{g}}$.

If we try to quantize the loop naively by viewing $J^a(z)$ as elements of $\hat{\textfrak{g}}$, we quickly run into difficulties. We are faced with ordering ambiguities, and infinite quantities appear from commutators when the ordered exponential is represented by a sum of integrals. In \cite{BG}, the authors were able to regularize the operators $J^a$, give an ordering prescription and add counterterms to finally get a finite result up to order four in $\frac{1}{k}$. The result was compatible with the intuition coming from quantum monodromy computations (\cite{BG}, section 3). This procedure, however, is only perturbative, and the computations are rather heavy already for the lowest orders. Hence, a prescription which would yield exact results would be desirable. The mathematical tools reviewed below will allow us to propose one.

\section{Central operators in Kac-Moody algebras}

\label{Kac}

In this section we review results from \cite{Kac}. We will show how to build central elements 
in some appropriate completion of the enveloping algebra of a Kac-Moody algebra, and see that these 
operators are naturally associated to complex functions of the weights. Note that similar results
for the Virasoro algebra appeared previously in \cite{FF}.

\subsection{Some preliminaries and the Shapovalov form}

We start with an affine Kac-Moody algebra\footnote{Actually the theorem proved in \cite{Kac} is valid for 
any Kac-Moody algebra.} $\hat{\textfrak{g}}$ with Cartan subalgebra $\hat{\textfrak{h}}$, Killing form $(.,.)$ and simple roots $\alpha_i \in \hat{\textfrak{h}}^*$. We denote by $\Delta$ the set of all roots, $Q = \sum_i \mathbbm{Z} \alpha_i$ the root lattice, $\Delta_+$ the set of positive roots and $Q_+  = \sum_i \mathbbm{N} \alpha_i$ ($0\in\mathbbm{N}$ in our conventions). For any element $\beta \in Q_+$, let $\beta = \sum_i k_i \alpha_i$ be its decomposition on simple roots and define $\mbox{deg}(\beta) = \sum_i k_i$. The Weyl vector $\hat{\rho}$ is defined as the sum of the fundamental weights.

We can choose Chevalley generators $\{e_i\}$ and $\{f_i\}$ that generate $\hat{\textfrak{g}}$ together with elements of $\hat{\textfrak{h}}$. They satisfy Serre relations and the following commutation identities :
$$
[e_i,f_j] = \delta_{ij}(\alpha^\vee_i)^\ast \;,\;\;\;\;\; \alpha^\vee = \frac{2\alpha}{(\alpha,\alpha)}\;,
$$
$$
[h, e_i] = \alpha_i(h) e_i \;,\;\;\;\;\; [h, f_i] = - \alpha_i(h) f_i\;,
$$
for $h \in \hat{\textfrak{h}}$. We call $\textfrak{n}_+$ and $\textfrak{n}_-$ the subalgebras generated by $\{e_i\}$ and $\{f_i\}$, respectively, and $\hat{\textfrak{g}}$ decomposes as : $\textfrak{n}_+ \oplus \hat{\textfrak{h}} \oplus \textfrak{n}_-$. Note that under the adjoint action of $\hat{\textfrak{h}}$, $U(\textfrak{n}_-)$ is decomposed into weight spaces $U(\textfrak{n}_-)_{-\beta}$ with $\beta \in Q_+$. There is an involutive anti-automorphism $\sigma$ exchanging $\textfrak{n}_+$ and $\textfrak{n}_-$ :
$$
\sigma(e_i) = f_i \;,\;\;\;\;\ \sigma(f_i) = e_i \;,\;\;\;\;\ \sigma|_\textfrak{h} = \mbox{id}_\textfrak{h} \;.
$$
It naturally extends to an anti-automorphism of the enveloping algebra $U(\hat{\textfrak{g}})$.

$\tilde{\Delta}_+$ will be the set of positive roots \emph{with} their multiplicities. Every positive root appears once in this set, except the imaginary root, with appears $r$ times, with $r$ the rank of the horizontal subalgebra. We choose some order on this set and denote the associated lowering operators by $F_1$, $F_2$, ... $\in \textfrak{n}_-$. For any $\beta$ belonging to the positive root lattice $Q_+$, we will call $\mathcal{P}_\beta$ the set of all maps :
$$
\begin{array}{ccccccc}
k&:& \tilde{\Delta}_+ & \rightarrow & \mathbbm{N}&& \\
&& F_i & \mapsto & k_i & \mbox{such that} & \beta = \sum_{i = 0}^{|\tilde{\Delta}_+|} k_i \alpha_i \;,
\end{array}
$$
where $\alpha_i$ is the positive root corresponding to the lowering operator $F_i$, and let $\mathcal{P} = \cup_{\beta\in Q_+} \mathcal{P}_\beta$. For any $k \in \mathcal{P}_\beta$, we define $|k| = \mbox{deg}(\beta)$ and :
$$
F^k = (F_1)^{k_1}(F_2)^{k_2}\ldots \in U(\textfrak{n}_-) \;.
$$
By the Poincaré-Birkhoff-Witt theorem, the set of all $F^k$ such that $k$ belongs to $\mathcal{P}_\beta$ form a basis of $U(\textfrak{n}_-)_{-\beta}$. We define a basis in $U(\textfrak{n}_+)$ : $E^k = \sigma(F^k)$, for $k \in \mathcal{P}$.

Let us choose $k,m \in \mathcal{P}_\beta$. Consider the free highest weight module $M_\lambda$ (Verma module) with highest weight $\lambda$ and highest vector $v_\lambda$. The Shapovalov form $B^\lambda_{\beta} : U(\textfrak{n}_+) \times U(\textfrak{n}_-) \rightarrow \mathbb{C}$ is defined by :
$$
E^kF^m v_\lambda = B^\lambda_{\beta}(E^k,F^m)v_\lambda \;.
$$

\subsection{A generalization of the enveloping algebra}

As it contains only polynomials in generators of $\hat{\textfrak{g}}$, the enveloping algebra $U(\hat{\textfrak{g}})$ is not large enough to contain central operators corresponding to the classical observables we aim to quantize (eg. Wilson loops). Therefore we need a suitable extension of the enveloping algebra $U(\hat{\textfrak{g}})$.

We first define affine linear functions on the weight space $\hat{\textfrak{h}}^*$ : $T_\beta(\lambda) = 2(\lambda + \hat{\rho},\beta) - (\beta,\beta)$, $\beta \in Q_+$. These functions are useful because of the following property \nolinebreak: if $T_{n\alpha}(\lambda) = 0$, for $n$ a positive integer, then the module $M_\lambda$ has a submodule isomorphic to $M_{\lambda - n\alpha}$. We name $L$ the union of all the hyperplanes of the form $T_{n\alpha}(\lambda + \gamma) = 0$, for $\alpha \in \Delta_+$ and $\gamma \in Q$. $\mathcal{F}$ will denote the algebra of holomorphic complex valued functions defined on $\hat{\textfrak{h}}^*\backslash L$. Note the canonical embedding $j : \hat{\textfrak{h}} \hookrightarrow \mathcal{F}$.


Let $U_\mathcal{F}(\hat{\textfrak{g}})$ be tensor product $U(\hat{\textfrak{g}})\otimes \mathcal{F}$ quotiented by the relations :
$$
h - j(h) = 0 \;,\;\;\;\;\; \phi g^\alpha = g^\alpha s_\alpha(\phi)\;,
$$
where $h \in \hat{\textfrak{h}}$, $\phi \in \mathcal{F}$, $g^\alpha$ a generator in the root space $\hat{\textfrak{g}}_\alpha$ and $s_\alpha$ acts as a shift on elements $\mathcal{F}$ : $s_\alpha(\phi)(\lambda) = \phi(\lambda + \alpha)$. Elements of $U_\mathcal{F}(\hat{\textfrak{g}})$ are finite sums of the form $\sum_{k,m\in \mathcal{P}}F^k\phi_{km} E^m$, $\phi_{km} \in \mathcal{F}$.
Because of the embedding $S(\hat{\textfrak{h}}) \subset \mathcal{F}$, we have $U(\hat{\textfrak{g}}) \subset U_\mathcal{F}(\hat{\textfrak{g}})$. $U_\mathcal{F}(\hat{\textfrak{g}})$ becomes a graded algebra if we define $\mbox{deg}(e_i) = -\mbox{deg}(f_i) = 1$.  

We further extend it by considering $\hat{U}_\mathcal{F}(\hat{\textfrak{g}}) = \bigoplus_j \hat{U}_\mathcal{F}(\hat{\textfrak{g}})_j$, where $\hat{U}_\mathcal{F}(\hat{\textfrak{g}})_j$ is the direct product of the subspaces of the form $F^k\mathcal{F}E^m$, with $|m| - |k| = j$. This amounts to extending $U_\mathcal{F}(\hat{\textfrak{g}})$ so that it contains infinite sums of the form $\sum_{k,m \in \mathcal{P}}F^k\phi_{km}E^m$ with $||k|-|m||$ bounded. Using the latter boundedness condition, it can be shown that the product is well-defined on $\hat{U}_\mathcal{F}(\hat{\textfrak{g}})$, in the sense that when multiplying two such series, only a finite number of terms contribute to a given grade. Hence $\hat{U}_\mathcal{F}(\hat{\textfrak{g}})$ keeps the structure of an algebra. Thanks to the same condition, it still has a well-defined action on any $\hat{\textfrak{g}}$-module with highest weight in $\hat{\textfrak{h}}^*\backslash L$. Indeed, when acting on a vector at a given grade, only a finite number of terms in the sum are non-vanishing.

The same construction can be applied to the symmetric algebra $S(\hat{\textfrak{g}})$ \nolinebreak: define $S_\mathcal{F}(\hat{\textfrak{g}})$ as the tensor product $S(\hat{\textfrak{g}}) \otimes \mathcal{F}$ quotiented by :
$$
h - \phi(h) = 0 \;,\;\;\;\;\; \phi g^\alpha = g^\alpha \phi \;.
$$
Again, a generic element of $S_\mathcal{F}(\hat{\textfrak{g}})$ is a finite sum of the form $\sum_{k,m\in \mathcal{P}}F^k\phi_{km} E^m$. Completing it to include infinite sums such that $||k|-|m||$ is bounded yields the completion $\hat{S}_\mathcal{F}(\hat{\textfrak{g}})$. 

The two algebras $\hat{S}_\mathcal{F}(\hat{\textfrak{g}})$ and $\hat{U}_\mathcal{F}(\hat{\textfrak{g}})$ admit an action of $\hat{\textfrak{g}}$. On $\hat{U}_\mathcal{F}(\hat{\textfrak{g}})$, it is provided by the adjoint action of $\hat{\textfrak{g}} \subset \hat{U}_\mathcal{F}(\hat{\textfrak{g}})$. For $\hat{S}_\mathcal{F}(\hat{\textfrak{g}})$, note that $\hat{\textfrak{g}}$ acts on $S(\hat{\textfrak{g}})$ by the adjoint action. This action can be extended to $\hat{S}_\mathcal{F}(\hat{\textfrak{g}})$ by defining for $\phi \in \mathcal{F}$ :
$$
\mbox{Ad}_{g^\alpha} \phi = g^\alpha(-\alpha \cdot \nabla \phi) \;,
$$
where $\alpha \in \Delta$, $g^\alpha$ is a generator in the root space $\hat{\textfrak{g}}_\alpha$, and $\nabla$ is the gradient on $\hat{\textfrak{h}}$. 

We will use the notation $\hat{S}_\mathcal{F}(\hat{\textfrak{g}})^{\hat{\textfrak{g}}}$ and $\hat{U}_\mathcal{F}(\hat{\textfrak{g}})^{\hat{\textfrak{g}}}$ for the invariant subalgebras with respect to these actions. The center $Z_\mathcal{F}$ of $\hat{U}_\mathcal{F}(\hat{\textfrak{g}})$ coincide with $\hat{U}_\mathcal{F}(\hat{\textfrak{g}})^{\hat{\textfrak{g}}}$. Note finally that elements of $Z_\mathcal{F}$ have the form $\sum_{\beta \in Q_+} \sum_{k,m\in \mathcal{P_\beta}}F^k\phi_{km} E^m$. 

\subsection{Building central operators}

The main theorem of \cite{Kac} states that given any function $\psi$ in $\mathcal{F}$, there is a central element $z_\psi \in Z_\mathcal{F}$ with eigenvalues on the modules $M_\lambda$ ($\lambda \in \hat{\textfrak{h}}^*\backslash L$) are equal to $\psi(\lambda)$. These operators are built recursively, by a constructive procedure.

As above, let $\psi$ in $\mathcal{F}$ and $z_\psi = \sum_{\beta \in Q_+} \sum_{k,m\in \mathcal{P_\beta}} F^k\phi_{km} E^m$ where $\phi_{km}$ are some unknown elements of $\mathcal{F}$ to be determined. $\lambda$ will denote an arbitrary weight of $\hat{\textfrak{h}}^*\backslash L$. Letting $z_\psi$ acts on $v_\lambda$, the highest weight vector of $M_\lambda$ and requiring that $z_\psi$ acts by scalar multiplication by $\psi(\lambda)$, we get $\phi_{00} = \psi$.
We will now compute the functions $\phi_{km}$ with $k,m \in \mathcal{P}_\beta$ by induction on $\beta$.

For $\gamma \in Q_+$, let $G_\gamma$ the matrix of the operator $\sum_{k,m\in \mathcal{P_\gamma}} F^k\phi_{km} E^m$ on the weight space $(M_\lambda)_{\lambda - \beta}$, in the basis $F^s(v_\lambda)$, $s\in \mathcal{P}_\beta$. We suppose that all the $G_\gamma$ are known for $\gamma < \beta$, and we write $\Phi_\beta$ the matrix $(\Phi_\beta)_{km} = \phi_{km}$, $k,m\in\mathcal{P}_\beta$. From the definition of $B_\beta$ above, we have :
$$
G_\beta = \Phi_\beta B_\beta
$$
with a matrix product between $\Phi_\beta$ and $B_\beta$ assumed.

We can now express the condition that $z_\psi$ acts by scalar multiplication on $(M_\lambda)_{\lambda-\beta}$ :
\begin{equation}
\label{recrelk}
\Phi_\beta B_\beta + \sum_{\gamma < \beta}G_\gamma = \psi(\lambda)\mathbbm{1}_\beta \;,
\end{equation}
where $\mathbbm{1}_\beta$ is the identity operator on $(M_\lambda)_{\lambda-\beta}$. Since det$B_\beta \neq 0$ on modules $M_\lambda)$ with $\lambda \in \hat{\textfrak{h}}^*\backslash L$ (see \cite{Kac} for details), this equation can be solved for $\Phi_\beta$. Hence all the coefficients $\phi_{km}$ in the expansion of the operator $z_\psi$ can be computed by recursion, showing its existence.

As this may seem a bit abstract, let us compute the next coefficient after $\phi_{00}$, namely the one corresponding to the first root appearing in the ordering we chose. This is a simple root, so the set $\mathcal{P_\alpha}$ has only one element. Rewriting (\ref{recrelk}) we get :
$$
\phi_{\alpha\alpha} B^\lambda_\alpha(f^\alpha,e^\alpha) + \psi(\lambda + \alpha) = \psi(\lambda) \;,
$$
which yields :
$$
\phi_{\alpha\alpha} = \frac{\psi(\lambda) - \psi(\lambda + \alpha)}{(\alpha^\vee,\lambda)} \;.
$$
Already in this simple example, it can be seen why we restricted the discussion to weights outside $L$. If $(\alpha^\vee,\lambda) = 0$ we cannot solve the equation for $\phi_{\alpha\alpha}$.

\subsection{Central operators on integrable modules}

We are interested in letting central operators act on integrable highest weight modules which compose the state space of WZW models, but such weights always lie in $L$. Even if we start with a function well behaved around integral weights, the central operators will in general have divergences in its coefficients $\phi_{km}$, as shown in the simple example above. Such divergences arise because the Shapovalov form $B_\beta$ is not invertible on modules with highest weight in $L$. Hopefully, there is a simple condition on the function $\phi$ warranting that these singularities cancel.

Let $K$ be the set of weights with positive (but not necessarily integral) level. Note that the set $K$, as well as $-\hat{\rho} + K$, contains all the dominant integrable weights. In \cite{Kac} it is proved that $z_\psi$ can be extended on the set $-\hat{\rho} + K$ if and only if for any positive integer $n$ and any real positive root $\alpha$, $T_{n\alpha}(\lambda) = 0$ implies $\psi(\lambda) = \psi(\lambda - n\alpha)$. The necessity of this condition is easily understood if we recall that $T_{n\alpha}(\lambda) = 0$ implies that the free module $M_\lambda$ contains a submodule isomorphic to $M_{\lambda - n\alpha}$. As $z_\psi$ acts by scalar multiplication by $\psi(\lambda)$ on the former and by $\psi(\lambda - n\alpha)$ on the latter these two values have to be equal for $z_\psi$ to be well-defined on $M_\lambda$. For the proof of sufficiency, we refer the reader to \cite{Kac}.

There is a slightly stronger and more convenient condition : for $z_\psi$ to be well defined on $-\hat{\rho} + K$, it is sufficient that $\psi$ is invariant under the action of Weyl group shifted by $\hat{\rho}$, ie. $\psi(w(\lambda + \hat{\rho}) - \hat{\rho}) = \psi(\lambda)$ for any $w \in W$.

Remark that shifting a weight by the affine Weyl vector $\hat{\rho}$ changes its level : $k \mapsto k + h^\vee$, where $h^\vee$ is the dual Coxeter number of the horizontal subalgebra $\textfrak{g}$.

\vspace{.5cm}

In summary, following \cite{Kac}, we constructed an injective ring homomorphism $Kac : \mathcal{F}^{\hat{W}_s} \rightarrow \hat{U}_\mathcal{F}(\hat{\textfrak{g}})^{\hat{\textfrak{g}}}$ from functions on $\hat{\textfrak{h}}^*$ invariant under the shifted Weyl group $\hat{W}_s$ to central operators living in $\hat{U}_\mathcal{F}(\hat{\textfrak{g}})$.

\vspace{.5cm}

\section{Quantization of invariant observables}

\label{quant}

\label{renlopop}

Let us use the results above to solve the problem described in section \ref{SectWloop}. Consider a classical observable of the WZW model depending holomorphically on the current $J(z)$ only, for instance the Wilson loop (\ref{WLoop}). It can be expanded as a series in the Fourier modes $J^a_n$ of the current $J$, so belongs to $\hat{S}_\mathcal{F}(\hat{\textfrak{g}})$. Moreover, the action of the Poisson bracket $\{.,J_n^a\}$ on such an observable is given by the action of $\hat{\textfrak{g}}$ on $\hat{S}_\mathcal{F}(\hat{\textfrak{g}})$. Therefore classical observables invariant under the symmetries of the WZW model are elements of $\hat{S}_\mathcal{F}(\hat{\textfrak{g}})^{\hat{\textfrak{g}}}$.

We have the following diagram :
$$
{\large
\xymatrix{
\hat{S}_\mathcal{F}(\hat{\textfrak{g}})^{\hat{\textfrak{g}}} \ar[ddd]_{R} \ar@{.>}[rrr]^{\tilde{Q}} &&& \hat{U}_\mathcal{F}(\hat{\textfrak{g}})^{\hat{\textfrak{g}}}  \\ &&& \\ &&& \\
\mathcal{F}^{\hat{W}} \ar[rrr]^{Sh} &&& \mathcal{F}^{\hat{W}_s} \ar[uuu]_{Kac} \;.
}
}
$$

$R$ is the restriction of $\hat{S}_\mathcal{F}(\hat{\textfrak{g}})$ on $\mathcal{F}$, which can be informally described as "setting to zero" the generators belonging to $\hat{\textfrak{g}}\backslash \hat{\textfrak{h}}$. It maps any $\hat{\textfrak{g}}$ invariant element of $\hat{S}_\mathcal{F}(\hat{\textfrak{g}})$ to a $\hat{W}$-invariant element of $\mathcal{F}$, because of the very definition of the Weyl group as the group of inner automorphisms preserving the Cartan subalgebra.

$Sh$ is the "shift" map. Given any function $\phi \in \mathcal{F}$, it acts by $Sh(\phi)(\lambda) = \phi(\lambda + \hat{\rho})$, where $\hat{\rho}$ is the Weyl vector of $\hat{\textfrak{g}}$. It maps functions invariant under the affine Weyl group $\hat{W}$ to functions invariant under the shifted affine Weyl group $\hat{W}_s$. 

$Kac$ is the homomorphism constructed in \cite{Kac} and reviewed in the previous section.

Defining $\tilde{Q} = Kac \circ Sh \circ R$, we get a "quantization" ring homomorphism associating a central operator to any classical observable invariant under the WZW symmetries. Note that in the case of a finite dimensional Lie algebra, this homomorphism is actually an isomorphism, the Duflo isomorphism \cite{Duf}.

\vspace{.5cm}

We proceed now to the quantization of the Wilson loop (\ref{WLoop}), with $l = \frac{1}{k}$. We are interested in the spectrum of the corresponding central operator $z_\mu$, so it is sufficient to compute the element $w_{\mbox{\tiny ren}}^{(\mu)} \in \mathcal{F}^{\hat{W}_s}$ obtained by applying $Sh \circ R$. Computing explicitly the action of $Kac$ allow to recover order by order the perturbative expansion of $z_\mu$ in term of the modes of the current, but this is probably useless for most applications.

We first compute the action of $R$. Consider (\ref{WLoop}), and suppose that the functions $J^a(z)$ all vanish, except if the generator corresponding to the index $a$ lies in a Cartan subalgebra $\textfrak{h}$. In this case, the argument of the exponential lie in a commutative algebra, and the exponential becomes a usual one (we can forget about the cyclic ordering). We will now make indices $i,j,...$ run on the index of the generators of $\textfrak{h}$, while indices $a,b,...$ still run over the whole $\textfrak{g}$. By decomposing the functions $J^a(z)$ into Fourier components $J^a_n$ we get :
$$
w^{(\mu)}(\lambda,k) := R(W^{(\frac{1}{k},\mu)}(J)) = \mbox{Tr}_{R_\mu} \exp \left ( \frac{1}{k} J^i_0 t^i \right ) = \chi_\mu(\frac{-2\pi i}{k} \lambda) \;,
$$
where we set $\lambda = -\frac{1}{2\pi i}(J^i_0 t^i)^* \in \textfrak{h}^*$, and $\chi_\mu$ is the character of $\textfrak{g}$ associated with the representation with highest weight $\mu$, viewed as a function on $\hat{\textfrak{h}}^*$. Note that because of the dependence on $k$, it is really a function on $\hat{\textfrak{h}}^*$ rather than on $\textfrak{h}^*$.

The map $Sh$ then acts in an obvious way : the finite part of the weight get shifted by $\rho$, the Weyl vector of $\textfrak{g}$, and the level is changed as  $k \mapsto k + h^\vee$, where $h^\vee$ is the dual Coxeter number. We get :
$$
w_{\mbox{\tiny ren}}^{(\mu)}(\lambda,k) := Sh(w^{(\mu)}(\lambda,k)) = \chi_\mu \left ( \frac{-2\pi i}{k + h^\vee} (\lambda + \rho) \right ) \;,
$$
so the quantum central operator is given by :
$$
z_\mu := Kac(w_{\mbox{\tiny ren}}^{(\mu)}) \;.
$$
By definition, $z_\mu$ has eigenvalue $\chi_\mu \left ( \frac{-2\pi i}{k + h^\vee} (\lambda + \rho) \right ) = \frac{S_{\mu\lambda}}{S_{0\lambda}}$ on $H_\lambda \otimes H_{\lambda^*}$. (Here we used a famous identity between finite characters and the modular $S$ matrix of $\hat{\textfrak{g}}$). This spectrum is consistent with the perturbative result of \cite{BG}, and matches perfectly their guess about the non-perturbative result.

The central operators $\{z_\mu\}$ form a ring (more precisely a semi-ring, as there is no additive inverse) isomorphic to the representation ring of $\textfrak{g}$, simply because their eigenvalues are given by the finite characters $\chi_\mu$. However, their eigenvalues on integrable modules form one-dimensional representations of the WZW fusion ring, by an argument completely analogous to the proof of the Kac-Walton formula. (See for instance \cite{cft}, §16.2.1.)

\section{Boundary perturbations and Wilson loop operators}

\label{BPert&WLOp}

In this section, we prove that under open/closed string duality, certain boundary perturbation in the open string picture is dual to the action of a Wilson loop on the corresponding boundary state in the closed string picture.

\subsection{The boundary perturbation}

\label{SubSBPert}

To be more specific, consider a cylindrical worldsheet $\{(\sigma_1, \sigma_2) \in [0,L] \times \mathbb{R}\}/(\sigma_2 \sim \sigma_2 + T)$ with some prescribed maximally symmetric boundary conditions $B_\mu$ on the boundary $\sigma_1 = 0$, and $B_\nu$ on the boundary $\sigma_1 = L$. In the ``open string picture'', the time runs along the periodic direction $\sigma_2$ of the cylinder, and the corresponding amplitude, after integrating over the modular parameter of the cylinder, computes the partition function for open strings stretching between the D-branes $B_\mu$ and $B_\nu$. 

The state space of open strings is given by a direct sum of irreducible integrable modules of $\hat{\textfrak{g}}$, with multiplicities given by the fusion coefficients of the WZW model :
$$
\mathcal{H}^{\mbox{\tiny open}}_{\mu\nu} = \bigoplus_{\eta \in P^+_k} \mathcal{N}_{\mu\nu}^{\;\;\;\eta} H_\eta \;.
$$
Here and below, ``$\mathcal{N}_{\mu\nu}^{\;\;\;\eta} H_\eta$'' will mean the direct sum of $\mathcal{N}_{\mu\nu}^{\;\;\;\eta}$ copies of $H_\eta$.
Let $w = \sigma_1 + i \sigma_2$, and define the following coordinate suitable for radial quantization of the system : $z = -\exp(-\frac{\pi i}{L}w)$. In the latter coordinate system, the hamiltonian reads (see for instance \cite{Saleur:1998hq}) :
$$
H = \frac{\pi}{L}(L_0 -\frac{c}{24}) \;,
$$
where $L_0$ is zero mode of the Sugawara stress tensor. 

We consider now a stack of $n$ D-branes of type $B_\mu$ on the boundary $\sigma_1 = 0$. To account for strings stretching between any of the $n$ branes $B_\mu$ and the single brane $B_\nu$, the state space of open strings gets tensored by $\mathbb{C}^n$. 
\begin{equation}
\label{PertStSpaceOp}
\mathcal{H}^{\mbox{\tiny open}}_{n\mu,\nu} = \bigoplus_{\eta \in P^+_k} \mathcal{N}_{\mu\nu}^{\;\;\;\eta} H_\eta \otimes \mathbb{C}^n \;.
\end{equation}
We add a time-dependant hamiltonian density $J^a(w) A^a \delta(\sigma_1)$, which is supported on the $\sigma_1 = 0$ boundary. The perturbing hamiltonian can be expressed in term of the currents in the $z$ coordinates as follow \nolinebreak:
$$
l\Delta H(\sigma_2) = l\int_0^L d\sigma_1 J^a(w) A^a \delta(\sigma_1) = l\frac{\pi}{L} \sum_m J^a_m A^a z^m |_{z = -\exp(\frac{\pi}{L}\sigma_2)} \;,
$$
where $\{A^a\}$ is a set of (dim$\textfrak{g}$) $n\times n$ matrices acting on the factor $\mathbb{C}^n$ of the open string state space, and $l$ is, as above, a real coupling. The sum on the Lie algebra index $a$ will always be implicit.

In the original Kondo problem (where $G = SU(2)$), this perturbation describes the interaction of $k$ electrons channels with an impurity of spin $\frac{n+1}{2}$, $J$ being the spin current associated with the electrons \cite{AFF}. In the string theory interpretation which we will adopt in section \ref{app}, it amounts to turning on a constant gauge field on a set of $n$ D0-branes \cite{ARS}.

\vspace{.5cm}

We will now specialize to the case when $l = \frac{1}{k+h^\vee}$ and $\{A^a\}$ forms an irreducible representation of $\textfrak{g}$. We will show that the perturbed theory is conformal and admits the same Kac-Moody spectrum-generating algebra as the original theory.

\subsection{Identification of the perturbed theory}

The crucial step to proceed is known for some time already (see \cite{AFF} for instance) : for the special value of the coupling chosen above, the perturbed hamiltonian is quadratic in new Kac-Moody currents $\{\tilde{J}^a_n\}$, defined by $\tilde{J}^a_n = J^a_n + A^a (-\exp(\frac{\pi}{L}\sigma_2))^n$. 

The \emph{evaluation} module $L_\lambda(z)$ for a Kac-Moody algebra $\hat{\textfrak{g}}$ associated with a finite irreducible $\textfrak{g}$-module $L_\lambda$ with highest weight $\lambda$ is defined by :
$$
J^a_n \mapsto z^n A^a \;\;\;\;\;\;\; K \mapsto 0 \;,
$$
where $A$ is the representation of $\textfrak{g}$ on $L_\mu$, $z$ is a non-zero complex number and $K$ is the central generator of $\hat{\textfrak{g}}$.

We now see that when the matrices $\{A_a\}$ form a $\textfrak{g}$-representation of highest weight $\lambda$, the action of the currents $\{\tilde{J}^a_n\}$ on $H_\eta \otimes \mathbb{C}^n$ form a representation of $\hat{\textfrak{g}}$. This representation is the tensor product of the representation with highest weight $\eta$ with the evaluation representation with highest weight $\lambda$ and $z = -\exp(\frac{\pi}{L}\sigma_2)$. One can also check directly that $\{\tilde{J}^a_n\}$ do satisfy the commutation relations for Kac-Moody currents.

Then, the hamiltonian can be rewritten :
\begin{align*}
H_{\mbox{\tiny pert}} &= H + \frac{1}{k + h^\vee} \Delta H = \\%
&= \frac{\pi}{L} \left ( \frac{1}{2(k + h^\vee)} \sum_m \left ( :J^a_m J^a_{-m}: + 2J^a_m A^a (-\exp(\frac{\pi}{L}\sigma_2))^m \right ) - \frac{c}{24} \right ) = \\ %
&= \frac{\pi}{L} \left ( \frac{1}{2(k + h^\vee)} \sum_m \left ( :\tilde{J}^a_m \tilde{J}^a_{-m}: - A^aA^a  \right ) - \frac{c}{24} \right ) \;,
\end{align*}
where we ``completed the square'' to go from the second line to the last one. This hamiltonian is quadratic in the new currents $\{\tilde{J}^a_m\}$, and satisfies the commutation relation $[H_{\mbox{\tiny pert}},\tilde{J}^a_m] = -\frac{\pi}{L} m \tilde{J}^a_m$. These relations imply that any eigenvector of $H_{\mbox{\tiny pert}}$ should belong to a highest weight module for the currents if we expect the spectrum of $H_{\mbox{\tiny pert}}$ to be bounded from below. Indeed, a state of minimal energy must be annihilated by $\{\tilde{J}^a_m\}_{m > 0}$.

Note that the terms $- A^aA^a$ are equal to the Casimir operator of the irreducible $\textfrak{g}$-representation $\{A^a\}$, and hence they are multiples of the identity. They do not affect the eigenvectors of $H_{\mbox{\tiny pert}}$ and can be eliminated by a suitable redefinition of energy.

\vspace{.5cm}

Our task is now to diagonalize this hamiltonian by finding highest weight modules for the representation of $\hat{\textfrak{g}}$ generated by the currents $\{\tilde{J}^a_n\}$. It is rather easy to show that the modules $H_\eta \otimes L_\lambda(z)$ do not contain any highest weight state for $\{\tilde{J}^a_n\}$.  However we expect to find highest weight modules as linear subspaces of an appropriate completion of $H_\eta \otimes L_\lambda(z)$.

This situation is analogous in the case of a free boson $X$, and we recall some well-known facts about this theory that should make more clear the steps to be taken below. Denote by $\{a_n\}$ the Fourier modes of the current $\partial X$, $[a_n,a_m] = n \delta_{n+m}$, $F_p$, $p \in \mathbb{R}$ the Fock module generated by $\{a_n\}_{n<0}$ from a highest weight state $\ket{p}$ such that $a_n \ket{p} = 0$ for $n < 0$ and $a_0 \ket{p} = p \ket{p}$. This Fock module do not contain any highest weight state for the perturbed currents $a_n \mapsto a_n + l \cdot 1$, $l \in \mathbb{R}$. However, there exists a highest weight state living in the completion given by the full linear span $\bar{F}_p$ of $F_p$, namely : $\ket{p_{\mbox{\tiny pert}}} = \exp \left ( -l \sum_{n<0} \frac{1}{n} a_n \right ) \ket{p}$. This ``coherent state'' is not normalizable with respect to the original norm on $F_p$, so it really lies in $\bar{F}_p \backslash F_p$. But the new Fock module $F^{\mbox{\tiny pert}}_p$ generated from $\ket{p_{\mbox{\tiny pert}}}$ by the action of the negative modes of perturbed currents is isomorphic to $F_{p+l}$. This isomorphism can therefore be used to define a hermitian product on $F^{\mbox{\tiny pert}}_p$, turning it into a Hilbert space. The latter is the physical space of the perturbed theory, which is nothing but another free boson with zero mode shifted by $l$. 

We will repeat this scenario to identify the Hilbert space of the perturbed theory. To this aim, we will use the vertex operators associated to primary WZW fields. 

A \emph{vertex operator} is a map :
$$
\Phi_\lambda^u(z) : H_\sigma \rightarrow  \hat{H}_\eta \;,
$$
where $\hat{H}_\eta$ denotes the completion given by the full linear span of the basis vectors of $H_\eta$ (recall that the latter is the Hilbert space consisting of vectors of finite norm). It is parametrized by $u\in L_\lambda$ and $z\in \mathbb{C}^\ast$. Vertex operators satisfy the gauge relations :
$$
[J^a_n, \Phi_\lambda^u(z)] = z^n \Phi_\lambda^{A^au}(z) \;,
$$
$\{A^a\}$ forming the representation of $\textfrak{g}$ on $L_\lambda$.
For fixed $u$ and $z$, and for each triple $(\sigma, \eta, \lambda)$ of integrable weights there exists a $\mathcal{N}_{\eta\lambda}^{\;\;\;\sigma}$-dimensional linear space of such operators (see \cite{KniZam}, or \cite{TsuKan} for a rigorous analysis in the case $G = SU(2)$). We choose a basis $\{\Phi_\lambda^u(z,m) \}$ in this $\mathcal{N}_{\eta\lambda}^{\;\;\;\sigma}$-dimensional linear space, with $m = 1,...,\mathcal{N}_{\eta\lambda}^{\;\;\;\sigma}$.

Let $\{u_i\}$ be an orthonormal basis of $L_\lambda$ and define $\Phi_\lambda(z,m) = \sum_{i} \Phi^{u_i}_\lambda(z,m) \otimes u_i$. The gauge relations above imply that $\Phi_\lambda(z,m)$ intertwines the action of $\hat{\textfrak{g}}$ on the highest weight module $H_\sigma$ and its action on the tensor product $\hat{H}_\eta \otimes L_\lambda(z)$. For each $\sigma \in P^+_k$, we deduce the following inclusion of $\hat{\textfrak{g}}$-modules :
$$
\bigoplus_{m=1}^{\mathcal{N}_{\eta\lambda}^{\;\;\;\sigma}} \Phi_\lambda(z,m) \left ( H_\sigma \right ) \subset \hat{H}_\eta \otimes L_\lambda(z) \;.
$$
Tying these maps together, we get an injective homomorphism of $\hat{\textfrak{g}}$-modules :
\begin{equation}
\label{InjMod}
\bigoplus_{\sigma \in P^+_k} \mathcal{N}_{\eta\lambda}^{\;\;\;\sigma} H_\sigma \stackrel{i}{\rightarrow} \hat{H}_\eta \otimes L_\lambda(z) \;,
\end{equation}
where ``$\mathcal{N}_{\eta\lambda}^{\;\;\;\sigma} H_\sigma$'' means the direct sum of $\mathcal{N}_{\eta\lambda}^{\;\;\;\sigma}$ copies of $H_\sigma$.

Any highest weight submodule contained in $\hat{H}_\eta \otimes L_\lambda(z)$ lies in the image of the sum of modules of the left-hand side, because we summed over a maximal set of linearly independent vertex operators. The isomorphism between $\bigoplus_{\sigma \in P^+_k} \mathcal{N}_{\eta\lambda}^{\;\;\;\sigma} H_\sigma$ and its image in $\hat{H}_\eta \otimes L_\lambda(z)$ provides $i \left ( \bigoplus_{\sigma \in P^+_k} \mathcal{N}_{\eta\lambda}^{\;\;\;\sigma} H_\sigma \right )$ with a hermitian form which shows that it is a Hilbert subspace of $\hat{H}_\eta \otimes L_\lambda(z)$. 

\vspace{.5cm}

Returning to our physical problem, we recognize in $\hat{H}_\eta \otimes L_\lambda(z)$ the typical term in the direct sum (\ref{PertStSpaceOp}) forming the perturbed state space of open strings, when the matrices $A^a$ form a irreducible representation of $\textfrak{g}$ of highest weight $\lambda$. We therefore showed that the physical state space of the perturbed theory of open strings stretching between branes $B_\mu$ and $B_\nu$ is given by the following direct sum of highest weight $\hat{\textfrak{g}}$-modules :
$$
\mathcal{H}'^{\mbox{\tiny open}}_{n\mu,\nu} = \bigoplus_{\eta,\sigma \in P^+_k} \mathcal{N}_{\mu\nu}^{\;\;\;\eta} \mathcal{N}_{\eta\lambda}^{\;\;\;\sigma} H_\sigma \;.
$$

\subsection{Turning to the closed string picture}

\label{TurnClosedString}

We will identify the special open string boundary perturbations considered above to Wilson loop operators in the closed string picture.

Now that we identified the state space of the perturbed theory and diagonalized its hamiltonian, we can easily compute its partition function :
$$
Z^{\mbox{\tiny open}}_{n\mu,\nu} = \sum_{\eta,\sigma \in P^+_k} \mathcal{N}_{\mu\nu}^{\;\;\;\eta} \mathcal{N}_{\eta\lambda}^{\;\;\;\sigma} \chi_\sigma (q) \;,
$$
where $\chi_\sigma (q) = \mbox{Tr}_{H_\sigma}  \left ( q^{\frac{1}{2}(L_0 - \frac{c}{12})} \right )$ are the specialized Kac-Moody characters, $q= e^{\frac{\pi iT}{L}}$ and $L_0 = \frac{1}{2(k+h^\vee)} \sum_n :J_n^a J_{-n}^a:$.

This partition function can be interpreted as an amplitude in the ``closed string picture'' by performing a modular transformation $q \mapsto \tilde{q} = e^{\frac{\pi iL}{T}}$ under which the characters transform as $\chi_\sigma (q) = \sum_{\xi \in P^+_k} S_{\sigma\xi} \chi_\xi (\tilde{q})$. We get :
\begin{align*}%
Z^{\mbox{\tiny open}}_{n\mu,\nu} &= \sum_{\eta,\sigma, \xi \in P^+_k} \mathcal{N}_{\mu\nu}^{\;\;\;\eta} \mathcal{N}_{\eta\lambda}^{\;\;\;\sigma} S_{\sigma\xi} \chi_\xi (\tilde{q}) = \sum_{\xi \in P^+_k} \frac{S_{\mu\xi}S_{\nu\xi}}{S_{0\xi}} \frac{S_{\lambda\xi}}{S_{0\xi}} \chi_\xi (\tilde{q}) = \\%
&= \bra{B_\nu} \tilde{q}^{\frac{1}{2}(L_0 + \bar{L}_0 - \frac{c}{12})} z_\lambda \ket{B_\mu} \;,%
\end{align*}
where we used Verlinde's formula $\mathcal{N}_{\mu\nu}^{\;\;\;\eta} = \sum_{\sigma \in P^+_k} \frac{S_{\mu\sigma}S_{\nu\sigma}\bar{S}_{\eta\sigma}}{S_{0\sigma}}$ and the elementary properties of the modular $S$ matrix.

\vspace{.5cm}

The partition function of the perturbed open string theory therefore coincide with the closed string picture amplitude between the boundary states $\bra{B_\nu}$ and $z_\lambda \ket{B_\mu}$. The complicated procedure we followed to identify the perturbed open string theory simply amounts in the closed string picture to the action of the Wilson loop operator corresponding to the perturbation on the boundary state.

Let us remark that this demonstration can be easily generalized to the case of symmetry breaking perturbations to be considered below in section \ref{app}. 

\subsection{Boundary RG flows}

We showed above how to integrate the finite boundary perturbation at the special coupling value $l = \frac{1}{k+h^\vee}$ and when the perturbing matrices form a representation of the horizontal Lie algebra $\textfrak{g}$. We want to discuss here the relation to boundary renormalization group flows.

The renormalization group flow procedure goes as follows. First, the system is perturbed at the boundary with an infinitesimal coupling $l = \epsilon$. This breaks the Kac-Moody and conformal symmetries of the open string theory, but allows the use of perturbation theory. As the theory is not conformal anymore, a rescaling of the worldsheet acts non-trivially on the space of couplings and induces a renormalization group flow. An infinite rescaling should eventually push the theory to a fixed point of the flow, which should be a conformal field theory. In principle, this procedure can be applied to any perturbation, but the identification of the fixed point is difficult, and often its mere existence remains unproven. Hence for the aim of building new branes using the perturbation discussed above, our simple and rigorous procedure advantageously replace RG flow techniques.

However, the main interest of RG flows lies in their conjectured link with time evolution. It is generally believed that if two branes $B_1$ and $B_2$ are linked by a boundary RG flow, there is a physical process turning $B_1$ into $B_2$ (Though this is far from obvious there has been recently some progress \cite{Freedman:2005wx} to clarify these issues.) This would imply that branes linked by boundary flows carry the same conserved charges (Ramond-Ramond charges when the theory is supersymmetrized), which makes RG flow a useful tool for the study of brane charges (see \cite{Fredenhagen:2004xp}, \cite{Fredenhagen:2000ei}).

We proved above that there exists a fixed point of the RG flow at $l= \frac{1}{k+h^\vee}$ when $\{A^a\}$ forms a representation of the horizontal Lie algebra. The fixed point theory is not only conformal, but admit the same Kac-Moody algebra as a spectrum generating algebra. Then a natural question to ask is whether the finite perturbation we managed to integrate corresponds to the endpoint of the flow generated by the corresponding infinitesimal perturbation. In other words, is there a RG flow trajectory linking the boundary states $\ket{B}$ and $z_\mu \ket{B}$ ?
 
To argue why this is the case, we want first to insist on the universality property of Wilson operators, following \cite{BG}. These operators are built ``in the bulk'' (ie using only the bulk field $J$), without any reference to a boundary state. Therefore one can think about the RG flow as acting on the operator. Letting this operator act on a boundary state will induce a boundary RG flow on the state, but the flow on the operator is independent of the boundary condition.

A perturbative analysis (in $\frac{1}{k}$) of the RG action on Wilson operators with generic coupling $l$ has been carried out in \cite{BG}. In particular, the beta functions have been computed and from these results it seems clear that the only fixed points for such perturbation occur at $l=0$ or $l = \frac{1}{k+h^\vee}$ (see figure 2 in \cite{BG}), at least for sufficiently large levels. As the fixed point at $l=0$ is unstable, perturbing the open string theory infinitesimaly and carrying a RG flow will make the coupling grow until it reaches the stable fixed point $l = \frac{1}{k+h^\vee}$. One could argue that $l$ is not the only coupling that could be altered by the flow, actually there are dim$\textfrak{g}\,n^2$ parameters corresponding to the matrix elements of the matrices $A^a$. However the analysis of \cite{ARS} and \cite{Monnier:2005jt} (again in the large level limit) shows that irreducible representations are local minima of the effective action, hence the directions transverse to $l$ should be stable.

It seems very likely that the picture stays the same at low level, even if the lack of control on the non-conformal theories in the interval $0<l<\frac{1}{k+h^\vee}$ does not allow any rigorous statements for now.

Because it provides explicitly the spectrum of the Wilson operator at $l = \frac{1}{k+h^\vee}$, our procedure allows to determine the endpoint of the RG flow induced on boundary states, modulo the slight reserve formulated above.

Let us also note that the ring structure carried by Wilson operators confirms the conjectures of \cite{Fredenhagen:2000ei} concerning RG flows at low level and is compatible with K-theory computations \cite{Braun:2003rd}, as already noted in \cite{BG}. One can also remark that when acting with a Wilson operator on some boundary state, the ratio of the g-factor of the final and initial boundary state is given by the eigenvalue on the module with zero weight. Hence these operators also contain informations about the directions of the flows induced by the corresponding boundary perturbation.

\section{An application}

\subsection{The symmetry breaking ground state of WZW branes}

\label{app}

Now we use the formalism developed above to identify the symmetry breaking boundary states described in \cite{Monnier:2005jt}. There it was shown using the effective action of \cite{ARS} that such brane configurations have lower energy than the familiar maximally symmetric ones. Arguments showing that they should correspond to the brane ground state were given. However, they were pictured only as states resulting from a perturbation and a boundary state description was missing.

Let us consider the following situation. We start with a stack of $n$ D0-branes, $n|B_0\rangle$, and then apply a perturbation at the boundary of the form described in section \ref{SubSBPert}, $\{A^a\}$ being a priori a set of generic $n \times n$ matrices. Physically, this perturbation amounts to turning on a gauge field $A$ on the stack of D0-branes. As the D0-branes have (classically at least) no spatial extension, the components $A_a$ are really constant matrices. This perturbation is marginal only in the limit $k \rightarrow \infty$, so in general a RG flow is needed to drive the theory back to a conformal point. 

The effective action used in \cite{Monnier:2005jt} computes (in the limit $k \rightarrow \infty$, $\alpha' \rightarrow 0$) the logarithm of the g-factor \cite{AL} of the brane configuration resulting from the boundary RG flow, as a function of the boundary parameter $A$. Up to a global factor, it is given by :
\begin{equation}
  \label{effa}
	S(A) = \left (-\frac{1}{4} \sum_{a,b} \mbox{Tr}([A_a,A_b][A^a,A^b]) + \frac{1}{3} \sum_{a,b,c} f^{abc} \mbox{Tr}(A_a[A_b,A_c]) \right ) \;.
\end{equation}
One can use it to compute the energy of various brane configurations and test their stabilities. It is not difficult to prove that the maximally symmetric brane configurations (the ones which correspond to non-trivial representations of $\textfrak{g}$) have lower energy than the starting set of D0-brane. This shows its instability against condensation. 

Perhaps more surprisingly, it was shown \cite{Monnier:2005jt} that some states usually have much lower energy than maximally symmetric ones. These states arise when $A$ forms an irreducible representation of a su(2) subalgebra of minimal embedding index (which therefore equals 1). Minimal index subalgebras are obtained from the Dynkin diagram of $\textfrak{g}$ by deleting all but one node associated with a long root. For simply laced Lie algebras, they coincide with the regular su(2) subalgebras. From now on, a ``su(2) subalgebra'' will always be a su(2) subalgebra of minimal embedding index. After picking such a subalgebra, we can choose some basis $\{e^a\}$ of $\textfrak{g}$ orthogonal with respect to the Killing form, and assume that the subalgebra is generated by $\{e^1, e^2, e^3\}$. To get the minimum of the action we then set $A_a$ ($a = 1,2,3$) to some matrices representing irreducibly $su(2)$, and $A_a = 0$ ($a > 3$).
Clearly the perturbation operator partially breaks the $\hat{\textfrak{g}}$-symmetry of the model, and the corresponding boundary states were unknown.

We want to identify these states with the ones discovered in \cite{Quella:2002ct} and reviewed in section \ref{SymBreakBC}. Arguments developed in section \ref{BPert&WLOp} (or a slight generalization thereof, see section \ref{RGFlowsSymBreak} below)  show that the endpoint of the RG flow is given by the action of a Wilson operator on the boundary state. As the set of D0-branes we start with is maximally symmetric, and the resulting state is not, this operator cannot be central with respect to $\hat{\textfrak{g}}$. However, as a $\widehat{\mbox{su}}(2)$ symmetry is preserved at the boundary, we know that it belongs to the center of the $\widehat{\mbox{su}}(2)$ subalgebra of $\hat{\textfrak{g}}$ generated by the embedding su(2)$\subset \textfrak{g}$. Note that the level of the su(2)-subalgebra is the same as the level of $\hat{\textfrak{g}}$, because the embedding index is 1, so all the affine algebras to be considered below are at level $k$.

The classical Wilson operator is written :
$$
W^{(l, \sigma)} = \mbox{Tr P}\exp\oint l A_a J^a(z) dz \;,
$$
where $l$ is some real constant (that we will take to be equal to $\frac{1}{k}$), and $\sigma$ is the su$(2)$ weight of the representation $A$. As $A_a = 0$ for $a > 3$, this perturbation involves only currents associated with the $\widehat{\mbox{su}}(2)$ subalgebra. Hence we can treat it exactly as we did for maximally symmetric Wilson loops in section \ref{renlopop}. This yields the following spectrum \nolinebreak:
$$
w_{\mbox{\tiny ren}}^{(\sigma)}(\zeta, k) = \chi^{\mbox{\tiny su}(2)}_\sigma \left ( \frac{-2\pi i}{k + 2}(\zeta + \rho) \right ) = \frac{S^{\widehat{\mbox{\tiny su}}(2)}_{\sigma\zeta}}{S^{\widehat{\mbox{\tiny su}}(2)}_{0\zeta}} 
$$
on a $\widehat{\mbox{su}}(2)$-module of weight $\zeta$. In the formula above, the Weyl vector $\rho$ and the modular matrix $S^{\widehat{\mbox{\tiny su}}(2)}$ are those of $\widehat{\mbox{su}}(2)$, and we used $h^{\vee} = 2$ for su(2). Applying the map $Kac$, we get a Wilson loop operator $z^{\widehat{\mbox{\tiny su}}(2)}_\sigma$, central with respect to the $\widehat{\mbox{su}}(2)$ subalgebra. The chiral sector $H_\mu$ of the WZW model Hilbert space decomposes into (an infinite number of) $\widehat{\mbox{su}}(2)$-modules under the action of the subalgebra $\widehat{\mbox{su}}(2)$, and $z^{\widehat{\mbox{\tiny su}}(2)}_\sigma$ acts by scalar multiplication by $\frac{S^{\widehat{\mbox{\tiny su}}(2)}_{\sigma\zeta}}{S^{\widehat{\mbox{\tiny su}}(2)}_{0\zeta}}$ on submodules of $\widehat{\mbox{su}}(2)$-weight $\zeta$.

We are now ready to compute the boundary state we are looking for. The starting boundary state is a set of $n$ D0-branes, which is just the sum of all Ishibashi states, with weight $\sqrt{S_{0\lambda}}$ :
$$
n |B_{0}\rangle = n \sum_{\mu \in P^+_k} \sqrt{S_{0\mu}}|\mu \rangle \! \rangle = n \!\!\!\!\!\!\!\!\! \sum_{(\mu, \zeta) \in P_{G/SU(2)}} \!\!\!\!\!\!\!\!\! \sqrt{S_{0\mu}} |\mu, \zeta \rangle \! \rangle \;.
$$
In the second equality, we expressed the D0-brane state in term of the Ishibashi states of \cite{Quella:2002ct}. Now we apply the Wilson loop operator\footnote{Note that we can write such an equality because the eigenvalues of the Wilson loop operator coincide on submodules related by the coset identification, hence it really acts by scalar multiplication on the Ishibashi state $|\mu, \zeta \rangle \! \rangle$.} :
$$
|B_{\mbox{\tiny final}}\rangle = z^{\widehat{\mbox{\tiny su}}(2)}_\sigma |B_{0}\rangle = \!\!\!\!\!\!\!\!\! \sum_{(\mu, \zeta) \in P_{G/SU(2)}} \!\!\!\!\!\!\!\!\! \sqrt{S_{0\mu}} \; \frac{S^{\widehat{\mbox{\tiny su}}(2)}_{\sigma\zeta}}{S^{\widehat{\mbox{\tiny su}}(2)}_{0\zeta}} |\mu, \zeta \rangle \! \rangle \;.
$$
Comparing with (\ref{sbst}), we find that $|B_{\mbox{\tiny final}}\rangle = |B_{(0, \zeta)} \rangle$.

\vspace{.5cm}

Let us make some remarks :
\begin{itemize}
  \item As already mentioned above, in the $k \rightarrow \infty$ limit, computations using the effective action (\ref{effa}) gave good evidence that the states $|B_{(0, \zeta)} \rangle$ have the lowest g-factor \cite{Monnier:2005jt}. Numerical computations of the g-factors at finite level using the explicit form of the boundary states found above confirm this conjecture.
	\item Note that $z_\mu|B_{(0, \zeta)} \rangle = |B_{(\mu, \zeta)} \rangle$ where $z_\mu$ are the central operators of the preceding section. Therefore all the states decribed in \cite{Quella:2002ct} can be retrieved from the D0-brane state $|B_0 \rangle$ by perturbation. It is straightforward to see that the more general states described in \cite{Quella:2002ns} can also be retrieved in the same way. 
	\item Because they act by scalar multiplication on the modules of the symmetry algebra preserved by the branes, the Wilson loop operators always commute with the Virasoro generators. This allows for instance to obtain interesting relations between partition functions (see also \cite{Petkova:2000ip}). Let us denote by $Z_{(\nu,\sigma),(\mu,\zeta)}$ the partition function of closed strings propagating between the boundary states $\bra{B_{(\nu,\sigma)}}$ and $\ket{B_{(\mu,\zeta)}}$. We have \nolinebreak:
\begin{align*}
	Z_{(\nu,\sigma),(\mu,\zeta)} &=  \bra{B_{(\nu,\sigma)}} q^{\frac{1}{2}(L_0 + \bar{L}_0 + \frac{c}{12})} \ket{B_{(\mu,\zeta)}} = \\
	&= \bra{B_{(0,0)}} z_\nu z^{\widehat{\mbox{\tiny su}}(2)}_\sigma q^{\frac{1}{2}(L_0 + \bar{L}_0 + \frac{c}{12})}  z_\mu z^{\widehat{\mbox{\tiny su}}(2)}_\zeta \ket{B_{(0,0)}} = \\
  &=	\bra{B_{(0,0)}} q^{\frac{1}{2}(L_0 + \bar{L}_0 + \frac{c}{12})}  \mathcal{N}_{\mu\nu}^{\;\;\lambda}z_\lambda (\mathcal{N}^{\widehat{\mbox{\tiny su}}(2)})_{\sigma\zeta}^{\;\;\eta}z^{\widehat{\mbox{\tiny su}}(2)}_\zeta \ket{B_{(0,0)}} = \\
  &= \mathcal{N}_{\mu\nu}^{\;\;\lambda} (\mathcal{N}^{\widehat{\mbox{\tiny su}}(2)})_{\sigma\zeta}^{\;\;\eta} Z_{(0,0),(\lambda,\eta)} \;,
\end{align*}
where $\mathcal{N}_{\mu\nu}^{\;\;\lambda}$ and $(\mathcal{N}^{\widehat{\mbox{\tiny su}}(2)})_{\sigma\zeta}^{\;\;\eta}$ are the fusion rules of $\hat{\textfrak{g}}_k$ and $\widehat{\mbox{su}}(2)_k$. The product of Wilson loop operators was computed using the fact that $z_\mu$ is central and the preceding remark. Hence $Z_{(\nu,\sigma),(\mu,\zeta)}$ is a linear combination of $Z_{(0,0),(\lambda,\eta)}$ with positive integer coefficients. In particular Cardy's condition \cite{Cardy:1989ir} needs only to be checked on $Z_{(0,0),(\lambda,\eta)}$ (this was done for $Z_{(\nu,\sigma),(\mu,\zeta)}$ in \cite{Quella:2002ct}).
\end{itemize}

\subsection{Symmetry breaking perturbations and RG flows}

\label{RGFlowsSymBreak}

We can repeat the discussion of section \ref{BPert&WLOp} for symmetry breaking perturbations. By considering the non-maximally extended chiral algebra $\widehat{\mbox{su}}(2) \times \hat{\textfrak{c}}$ (recall that $\textfrak{c}$ is the chiral algebra of the coset model), we can go through the computations of sections \ref{SubSBPert} - \ref{TurnClosedString}, and we will only summarize the main points of the reasoning here.

Let $\hat{\textfrak{a}}$ be the minimal index $\widehat{\mbox{su}}(2)$ subalgebra to be preserved by the perturbation. We will consider the case when the boundary condition at $\sigma_1 = L$ is trivial and the one at $\sigma_1 = 0$ is a stack of $n$ identical symmetry breaking branes of the form (\ref{sbst}). The state space is given by \cite{Quella:2002ct} :
$$
\mathcal{H}^{\mbox{\tiny open}}_{n(\mu,\tilde{\mu}),(0,0)} = \!\!\! \bigoplus_{\tilde{\eta},\tilde{\sigma} \in P^+_k(\hat{\textfrak{a}})} \!\!\! (\mathcal{N}^{\hat{\textfrak{a}}})_{\tilde{\mu}\tilde{\eta}}^{\;\;\;\tilde{\sigma}} H^{\hat{\textfrak{a}}}_{\tilde{\eta}} \otimes H^{\hat{\textfrak{c}}}_{(\mu,\tilde{\sigma})} \otimes \mathbb{C}^n \;.
$$
(In this paragraph, $\widehat{\mbox{su}}(2)$-weights will be denoted by a tilde over them.) The open string hamiltonian is written :
\begin{equation}
\label{OpenHamSymBreak}
H = \frac{\pi}{L} \left ( L_0^{\hat{\textfrak{a}}} + L_0^{\hat{\textfrak{c}}} - \frac{c}{12} \right ) \;\;\;\;\;\;\; L_0^{\hat{\textfrak{a}}} = \frac{1}{k+2} \sum_{n \in \mathbb{Z}} : J^i_n J^i_{-n} : \;\;\; i = 1,2,3 \;.
\end{equation}
The open string perturbation is given by :
$$
l\Delta H(\sigma_2) = l\frac{\pi}{L} \sum_m J^i_m A^i z^m |_{z = -\exp(\frac{\pi}{L}\sigma_2)} \;,
$$
where the matrices $\{A^i\}$ form an irreducible representation of su(2) of highest weight $\tilde{\lambda}$. At $l = \frac{1}{k+2}$, we can ``complete the square'' in the perturbed hamiltonian $H_{\mbox{\tiny pert}} = H + \frac{1}{k + 2} \Delta H$ to show that it is quadratic in the new currents $\tilde{J}^i_n = J^i_n + A^i (-\exp(\frac{\pi}{L}\sigma_2))^n$, which generate a copy of $\widehat{\mbox{su}}(2)$ at level $k$. We have then the commutation relations :
$$
[H_{\mbox{\tiny pert}},\tilde{J}^i_m] = -\frac{\pi}{L} m \tilde{J}^i_m \;\;\;\;\;\;\;\;\;\; [\tilde{J}^i_m,L_n^{\hat{\textfrak{c}}}] = 0 \;\;\;\;\;\;\;\;\;\; [H_{\mbox{\tiny pert}},L_n^{\hat{\textfrak{c}}}] = -\frac{\pi}{L}n L_n^{\hat{\textfrak{c}}} \;,
$$
showing that $\widehat{\mbox{su}}(2) \times \hat{\textfrak{c}}$ generates the spectrum of the theory. Note however that the generators in $\hat{\textfrak{g}}\backslash \hat{\textfrak{a}}$ do not have any simple commutation relation with the hamiltonian anymore : even if the initial boundary state was maximally symmetric ($\tilde{\mu} = 0$), the symmetry is now broken to $\widehat{\mbox{su}}(2) \times \hat{\textfrak{c}}$.

The reasoning using vertex operators (of $\widehat{\mbox{su}}(2)$) is unchanged and allows us to identify the state space of the perturbed theory as :
$$
\bigoplus_{\tilde{\eta},\tilde{\sigma},\tilde{\zeta} \in P^+_k(\hat{\textfrak{a}})} \!\!\! (\mathcal{N}^{\hat{\textfrak{a}}})_{\tilde{\mu}\tilde{\eta}}^{\;\;\;\tilde{\sigma}}
(\mathcal{N}^{\hat{\textfrak{a}}})_{\tilde{\lambda}\tilde{\eta}}^{\;\;\;\tilde{\zeta}} H^{\hat{\textfrak{a}}}_{\tilde{\zeta}} \otimes H^{\hat{\textfrak{c}}}_{(\mu,\tilde{\sigma})} \;.
$$
The partition function of the perturbed theory is then :
\begin{align*}
Z^{\mbox{\tiny open}}_{n(\mu,\tilde{\mu}),(0,0)} &= \!\!\! \sum_{\tilde{\eta},\tilde{\sigma},\tilde{\zeta} \in P^+_k(\hat{\textfrak{a}})} \!\!\! (\mathcal{N}^{\hat{\textfrak{a}}})_{\tilde{\mu}\tilde{\eta}}^{\;\;\;\tilde{\sigma}}
(\mathcal{N}^{\hat{\textfrak{a}}})_{\tilde{\lambda}\tilde{\eta}}^{\;\;\;\tilde{\zeta}} \chi^{\hat{\textfrak{a}}}_{\tilde{\zeta}}(q) \chi^{\hat{\textfrak{c}}}_{(\mu,\tilde{\sigma})}(q) \\
&= \bra{B_{(0,0)}} \tilde{q}^{\frac{1}{2}(L_0 + \bar{L}_0 - \frac{c}{12})} z_{\hat{\lambda}}^{\hat{\textfrak{a}}} \ket{B_{(\mu,\tilde{\mu})}} \;,
\end{align*}
where a lengthy but straightforward computation was omitted.

\vspace{.5cm}

Therefore we showed that when the matrices $\{A^a\}$ form an irreducible representation of a su(2) subalgebra of minimal index, the perturbation at $l = \frac{1}{k+2}$ can be integrated exactly, and the resulting theory is dual to a theory with a symmetry breaking Wilson operator inserted in the closed string picture. This shows the existence of a fixed point at which only part of the original Kac-Moody symmetry is restored. 

To gain more information about the induced RG flow, one has to study its action on the symmetry breaking Wilson operators at generic coupling $l$. But the latter are exactly the same as the central Wilson operators in the WZW model with $G = SU(2)$. Therefore all the conclusions of \cite{BG} and section \ref{BPert&WLOp} apply. We know the existence of a stable fixed point at $l = \frac{1}{k+2}$, and we guess that there is a RG flow connecting it to the identity operator, at least for large $k$, even if it seems difficult to check it rigorously.

Therefore we have strong indication that all the known branes with trivial gluing automorphism (see \cite{Birke:1999ik, Gaberdiel:2002qa}) are linked by boundary RG flows, in spite of them being maximally symmetric or not. This also gives support to the ``plausible claim'' of \cite{Gaberdiel:2004za} (see the end of section 2 there).

\section*{Outlook}

The results presented here heavily rely on the facts that the perturbation involves no other field than the currents generating the symmetries of the model, and that the classical Wilson loop is an observable with some strong invariance properties. We are therefore unsure to which extent they could be generalized to other types of perturbations and conformal field theories.

However, as the example treated above should have demonstrated, modeling of the boundary renormalization group flow by central operators acting on the boundary states provides simple and powerful tool to identify the targets of the flows. Moreover, these exact results allow to build branes easily by mean of finite boundary perturbation. We hope to show in a future work how these techniques can be used to yield genuinely new branes, in particular branes preserving only the conformal symmetry of the model.

\section*{\normalsize Acknowlegments}

We would like to thank Anna Lachowska for discussions in the early stages of this project. This work was supported in parts by the Swiss National Fund for Scientific Research.

\end{document}